# The innermost regions of relativistic jets and their magnetic fields in radio-loud Active Galactic Nuclei

White Paper in Support of the Mission Concept of the Large Observatory for X-ray Timing

Authors

I. Donnarumma[1], I. Agudo[2], L. Costamante[3], F. D'Ammando[1,4], G. Giovannini[1,4], P. Giommi[5], M. Giroletti[1], P. Grandi[1], S. G. Jorstad[7,8], A. P. Marscher[7], M. Orienti[1], L. Pacciani[1], T. Savolainen[6,9], A. Stamerra[1,10], F. Tavecchio[1], E. Torresi[1], A. Tramacere[11], S. Turriziani[12], S. Vercellone[1], A. Zech[13]

[1] INAF, Italy
[2] Instituto de Astrofísica de Andalucía, Granada, Spain
[3] University of Perugia, Perugia, Italy
[4] Universitá degli Studi di Bologna, Bologna, Italy
[5] ASI Science Data Center (ASDC), Frascati, Italy
[6] Aalto University Metsähovi Radio Observatory, Metsähovintie 114, 02540 Kylmälä, Finland
[7] Institute for Astrophysical Research, Boston University, 725 Commonwealth Avenue, Boston, MA 02215, USA
[8] Astronomical Institute, St. Petersburg State University, Universitetskij Pr. 28, Petrodvorets, 198504 St. Petersburg, Russia
[9] Max Planck Institut für Radioastronomie, Bonn, Germany
[10] Scuola Normale Superiore, Piazza dei Cavalieri, 7, 56126 Pisa, Italy
[11] ISDC, University of Geneva, Switzerland
[12] University of Tor Vergata, Rome, Italy
[13] LUTH, Université Paris 7, Paris, France

## Preamble

The Large Observatory for X-ray Timing, *LOFT*, is designed to perform fast X-ray timing and spectroscopy with uniquely large throughput (Feroci et al., 2014). *LOFT* focuses on two fundamental questions of ESA's Cosmic Vision Theme "Matter under extreme conditions": what is the equation of state of ultra-dense matter in neutron stars? Does matter orbiting close to the event horizon follow the predictions of general relativity? These goals are elaborated in the mission Yellow Book (`http://sci.esa.int/loft/53447-loft-yellow-book/`) describing the *LOFT* mission as proposed in M3, which closely resembles the *LOFT* mission now being proposed for M4.

The extensive assessment study of *LOFT* as ESA's M3 mission candidate demonstrates the high level of maturity and the technical feasibility of the mission, as well as the scientific importance of its unique core science goals. For this reason, the *LOFT* development has been continued, aiming at the new M4 launch opportunity, for which the M3 science goals have been confirmed. The unprecedentedly large effective area, large grasp, and spectroscopic capabilities of *LOFT*'s instruments make the mission capable of state-of-the-art science not only for its core science case, but also for many other open questions in astrophysics.

*LOFT*'s primary instrument is the Large Area Detector (LAD), a 8.5 $m^2$ instrument operating in the 2–30 keV energy range, which will revolutionise studies of Galactic and extragalactic X-ray sources down to their fundamental time scales. The mission also features a Wide Field Monitor (WFM), which in the 2–50 keV range simultaneously observes more than a third of the sky at any time, detecting objects down to mCrab fluxes and providing data with excellent timing and spectral resolution. Additionally, the mission is equipped with an on-board alert system for the detection and rapid broadcasting to the ground of celestial bright and fast outbursts of X-rays (particularly, Gamma-ray Bursts).

This paper is one of twelve White Papers that illustrate the unique potential of *LOFT* as an X-ray observatory in a variety of astrophysical fields in addition to the core science.

## 1 Summary

The Large Area Detector (2–50 keV, LAD) aboard the *LOFT* mission thanks to its unprecentented area peaking at 8 keV ($\sim 8.5\,\mathrm{m}^2$) will have a great potential for the investigation of jet properties in radio-loud Active Galactic Nuclei (AGN). Active galactic nuclei (AGN) offer ideal laboratories to address several key problems of modern astrophysics. *LOFT* is planned in the same timeframe as other observatories that are opening up the AGN time domain on short timescales, e.g., the Cherenkov Telescope Array (CTA) at TeV energies and the Square Kilometer Array (SKA) and Atacama Large Millimeter Array (ALMA) in the radio and (sub-)mm bands, respectively. In this context, *LOFT*/LAD will be able to follow-up a large number of AGNs (up to 100 sources, see Table 1), while the LOFT/WFM can provide weekly monitoring (a few days for the brightest) thanks to its sky coverage.

- *Blazars* are the most variable sources over timescales that can be as short as minutes. This is the case of the High Energy Peaked BL Lac (HBL) objects showing strong variability in X-rays, which highly correlate with that of the TeV emission. The degree of this correlation is still debated, particularly when the flaring activity is followed down to very short time scales. We argue that the *LOFT* LAD, thanks to its timing capability, will allow us to detect the X-ray counterpart (2–50 keV) of the very fast variability observed at TeV energies, shedding light on the nature of the X-TeV connection. We also discuss the potential of *LOFT* in measuring the change in spectral curvature of the synchrotron spectra in HBLs which will make possible to directly study the mechanism of acceleration of highly energetic electrons. *LOFT* timing capability will be also promising in the study of Flat Spectrum Radio Quasars (FSRQs) with 2–10 keV flux $\gtrsim 2\times10^{-11}\,\mathrm{erg\,cm^{-2}\,s^{-1}}$. For this kind of objects, *LOFT* will probe the inverse Compton emission, whereas latest generation radio and (sub-)mm instruments will provide excellent timing capabilities to study short time-scale correlations with the synchrotron spectral domain. Constraints to the location of the high energy emission will be provided by: a) temporal investigation on second timescales; b) spectral trend investigation on minute timescales. This represents a further link with CTA because of the rapid (unexpected) TeV emission recently detected in some FSRQs. In this respect, the LOFT/WFM will provide not only an excellent trigger for both CTA and (sub-)mm observatories but also a long-term monitoring for a large sample of blazars.

- *Misaligned AGNs:* WFM long term monitoring (monthly timescale) of radio galaxies hosting efficient accretion disks (mainly FRIIs) will provide an optimal tool to investigate the disc-jet connection by triggering both radio and LAD X-ray follow-ups. *LOFT* will investigate the disc-jet connection in a tight link with X-ray binaries, which exhibit the same behavior.

  Recent detections of radio galaxies hosting inefficient accretion flows at high and very high energies (e.g., IC 310) will provide a further sample of radio-loud sources well suited for follow-up observations with LAD. Combining LAD and CTA simultaneous observations will be crucial in constraining the TeV emitting region, likely connected with the close surroundings of the central engine.

- *Radio-Loud Narrow-Line Seyfert 1* LAD observations will be very promising not only for studying the physical mechanisms responsible for the higher energy emission (in $\gamma$-rays or at TeV energies) and then the blazar-like behavior, but also for determining the role of its thermal component. This means to investigate the possible link between accretion and jet properties.

The LAD characteristics will therefore match very well with those of atmospheric Cherenkov telescopes such as the CTA, as well as the mm/radio observatories such as SKA and ALMA because all these observatories are timing explorers, characterized by huge collecting areas in the respective bands. With its huge collecting area peaking at 8–10 keV, the LAD will allow us to study more in depth the temporal and spectral evolution in the 2–50 keV band, for sources with fluxes above $2 \times 10^{-12}$ erg cm$^{-2}$ s$^{-1}$ (see Fig. 1). In the following sections, we will focus on the temporal and spectral investigation in all these radio-loud AGNs.

## 2 Blazars: the standard framework and its challenges

Blazars (Urry & Padovani, 1995) are among the most powerful persistent cosmic sources, able to release (apparent) luminosities exceeding $10^{48}$ erg s$^{-1}$ over the entire electromagnetic spectrum, from the radio to the very high energy $\gamma$-ray band. This intense non-thermal emission is produced within a relativistic (bulk Lorentz factor $\Gamma$ = 10–20) jet pointing toward the Earth. Relativistic effects, incorporated in the so-called relativistic Doppler factor $\delta$, lead to the beaming of the radiation within a narrow cone of semi aperture $\sim 1/\Gamma$ with strong amplification of the apparent luminosity and shortening of the variability timescales in the cases in which (as in blazars) the jet is closely aligned to the line of sight.

The spectral energy distribution (SED) of blazars displays two characteristic broad humps, whose peak frequencies appear to be anti-correlated with the luminosity (Fossati et al., 2011). The first component results from synchrotron emission of leptons in the jet. The origin of the second component, peaking in the $\gamma$-ray band, is more debated. In the most popular scenario it is attributed to the inverse Compton scattering between the synchrotron-emitting electrons and soft target photons, either the synchrotron photons themselves or those coming from the external environment. Alternatively, the high-energy bump could be the result of a population of relativistic hadrons in the jet, loosing energy either through synchrotron emission or photo-meson reactions with soft target photon fields (Böettcher et al., 2012).

Blazars occur in two flavors. Those showing optical broad emission lines as typically observed in quasars are called Flat Spectrum Radio Quasars (FSRQs). The weakest sources in general display rather weak or even absent emission lines and are collectively grouped in the BL Lac object class. The latter sources are those showing the most extreme variability (with variations as fast as few minutes) and the most energetic photons, with substantial emission above 100 GeV.

While the standard leptonic scenarios provide in general good descriptions of the blazar SEDs, they face difficulties when confronted with the most extreme HBLs (Cerruti et al., 2014), characterized by high-energy bumps peaking above ~1 TeV, and with intermediate objects (Böettcher et al., 2012), i.e. low-or intermediate-frequency peaked BL Lac objects (LBLs/IBLs).

The conventional scenario (e.g., Ghisellini & Tavecchio, 2009) foresees a single portion of the jet dominating the overall emission, at least during high activity states. However, this simple idea has been recently challenged by the observation of very fast (minutes) variability, which requires the existence of very compact emission sites. The most extreme example is PKS 2155−304, which experienced two exceptional flares in 2006, on top of which – with a recorded luminosity of the order of several $10^{47}$ erg s$^{-1}$ – events with rise times of ~100 s have been detected. Even assuming that these strong events originated in very compact emission regions, the physical conditions should be rather extreme and Lorentz factors of $\Gamma \sim 50$–100 seem unavoidable (Begelman et al., 2008). Physical mechanisms possibly triggering the formation of these compact emission regions include magnetic reconnection events (Giannios, 2013) or relativistic turbulence (Narayan & Piran, 2012; Marscher, 2014).

Alternatively, narrow beams of electrons can attain ultra-relativistic energies $\gamma \sim 10^6$ in the black hole vicinity through processes involving the BH magnetosphere (Rieger & Aharonian, 2008; Ghisellini et al., 2009). In this case, it is expected that the emission shows up only in high energy $\gamma$-ray region, giving no signals at lower frequencies (thus resembling the case of the so-called "orphan" flares; different from the simultaneous X-ray and TeV flaring observed from PKS 2155−304 in 2006).

Until now, ultra-fast variability has been recorded only in the $\gamma$-ray band, especially at TeV energies, at which the Cherenkov arrays are characterized by gigantic collection areas, required to probe such short timescales. Current instruments do not provide an analogous sensitivity in the key X-ray band, at which the low energy component of TeV emitting BL Lacs (the so-called HBLs) peaks and thus it is not possible to study potential ultra-fast variability events at these frequencies. Moreover, such events appear to be quite rare, with a duty cycle less than 1%. Therefore a monitoring of the sky is required to catch these events.

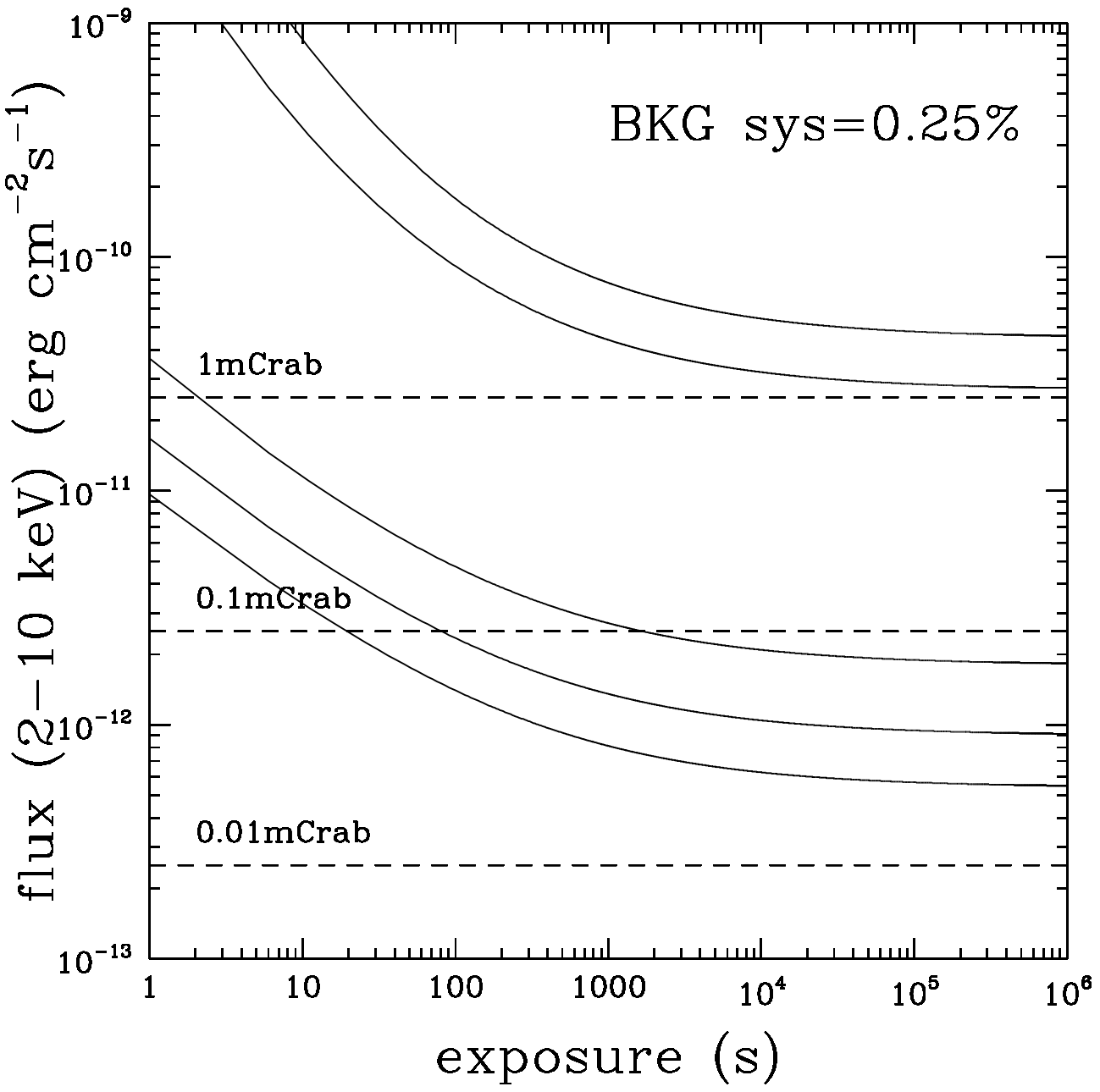

Figure 1: LAD sensitivity limits at different signal to noise ratio corresponding to a systematics of 0.25% on the background knowledge. From bottom to top: S/N ~ 3, 5, 10, 150, 250.

From this brief sketch it should be clear how the features of *LOFT* are ideal to address these key problems. In the following we discuss in detail some case studies.

## 2.1 High energy peaked BL Lac objects

### 2.1.1 X-ray/TeV connection

HBLs are good targets for the LAD as its energy band (2–50 keV) probes the synchrotron emission in the region above or close to the peak where the variability is expected to be higher.

The measurement of the TeV/X-ray lag also constrains the emission process: e.g., synchrotron self Compton (SSC) time dependent non-homogenous modeling predicts a TeV lag equal, approximately, to the light travel time across the emission region, whereas external Compton (EC), under the assumption the main source of variability is in the relativistic electron population of the jet, predicts almost simultaneity (Sokolov et al., 2004). Present data do not provide measurements of lags less than a few hundred seconds (Aharonian et al., 2009), while with LAD this timescale will be pushed to a few/tens of seconds. This means that for a 2–10 keV flux of the order of $10^{-10}$ erg cm$^{-2}$ s$^{-1}$, i.e., the typical flaring state for many HBLs and average state for some of them, the blazar flaring activity could be followed down to the decaying tail of the light curve with comparable bin time in X-ray and TeV energy bands (Sol et al., 2013). We show in Fig. 2 how LAD observations (red points) will detect a flare lasting ~60 s (which is one half of the shortest time bin explored so far in X-rays), shaped with a flux rise of 20% with respect to ~ $10^{-10}$ erg cm$^{-2}$ s$^{-1}$ over 120 s. We did the same simulations for both *Chandra* and *XMM-Newton*, adopting a time resolution of 2 s. It is worth noticing that the same statistical accuracy of the LAD can be achieved with *XMM* if a time bin an order of magnitude larger is considered. However, such a rebinning makes hard to define the flare rise time. Therefore, LAD observations will provide a real step-forward in the exploration of the X-ray and TeV correlation, possibly unveiling X-ray counterparts of TeV flaring structures.

These timing capabilities will provide a **unique** diagnostics to constrain the new emerging scenario of multi-zone (multiple electron populations) SSC modeling, as suggested by the degree of correlation observed between X-rays and TeV emissions during the flaring activity (Aharonian et al., 2009). The detection of flaring activity on very short timescale in X-ray and TeV will be crucial in the understanding of jet physics and its connection with the central engine: extreme variability implies very compact regions allowing us to investigate the properties of black holes and their surroundings.

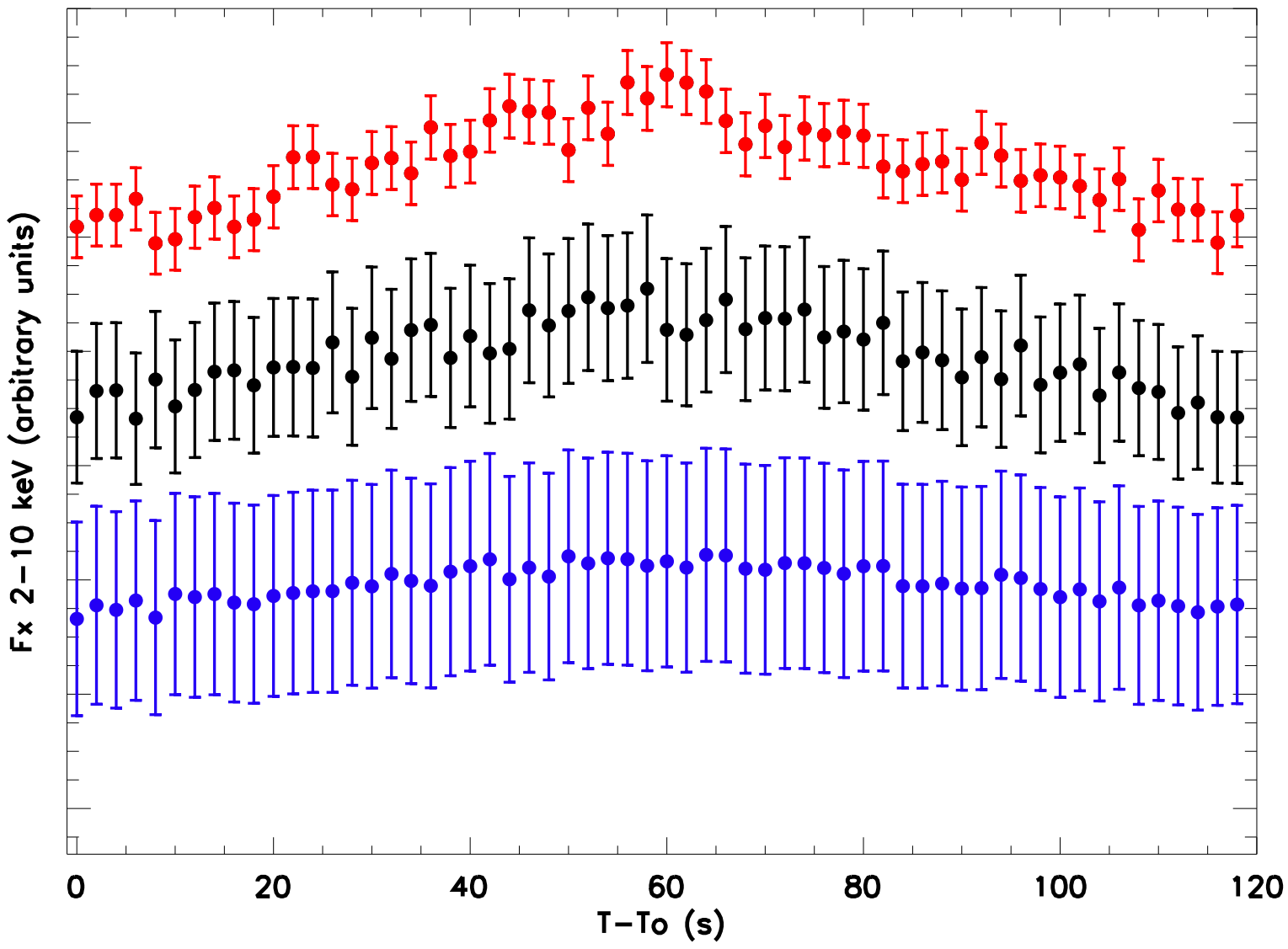

Figure 2: LAD (red), *XMM* (black, pn+MOS1,2) and *Chandra* (blue, ACIS-I) simulations of an X-ray enhancement of a 20% over a $\sim 10^{-10}\,\mathrm{erg\,cm^{-2}\,s^{-1}}$ in a rise time of one minute (this is one of the shortest timescale explored so far as in the case of PKS 2155−304). Each bin is 2 s wide.

While multi-waveband data on flares provide very significant constraints on our current blazar models, it is also important to investigate the temporal behaviour of blazars during quiescent states. In multi-zone models, where different emission regions could be responsible for the permanent high-energy emission on the one hand and for flares on the other, those regions might be identified by different variability patterns on short time-scales. Rapid variability in low states could be probed with *LOFT*/LAD for the most luminous HBLs, such as PKS 2155−304, Mrk 421 and Mrk 501, which will also figure among the most prominent sources for CTA.

For the most extreme HBLs with the highest peak energies, such as e.g. 1ES 0229+200 or 1ES 0347−121, although not the most luminous sources in X-rays, spectral data in the hard X-ray range, and especially information on temporal behaviour, will be accessible with *LOFT*/LAD and will help to better characterize the still poorly understood distribution of relativistic particles in those sources.

Moreover, the WFM will provide an excellent trigger for CTA (and HESS, MAGIC, VERITAS) observations as most of the targets suitable for TeV observation are relatively bright X-ray sources.

### 2.1.2 The acceleration mechanism

In the case of HBLs, a detailed measurement of the shape of the synchrotron spectral component from Optical/UV to hard X-ray frequencies provides physical information about the particle acceleration process in the jet since it directly traces the shape of the underlying particle distribution.

The broadband spectral distributions of several HBLs is well described by a log-parabolic fit, with the second-degree term measuring the curvature in the spectrum. This is thought to be a fingerprint of stochastic acceleration (Tramacere et al., 2007). In this scenario, the discrimination between a log-parabolic and power-law cut-off shape provides a powerful tool to disentangle acceleration dominated states, from states at the equilibrium, or very close to, giving solid constraints on the competition between acceleration and cooling times, and on the magnetic field intensity. We have simulated a pure log-parabolic spectrum peaking a 1.5 keV, resembling the *typical* spectrum of a HBL object in a quiescent state. We assumed a 2–10 keV flux of $\sim 10^{-10}\,\mathrm{erg\,cm^{-2}\,s^{-1}}$ over an integration time of 10 ks. In Fig. 3 we show the capability of the LAD to discriminate between a log-parabola (right panel) and a power-law cut-off model (left panel) with a high statistical significance. Thanks to the unprecedented *LOFT* large area peaking at 8 keV and its broad energy range this result provides a relevant improvement compared to the performance of focused X-ray instruments operating in the usual 0.2–10.0 keV range.

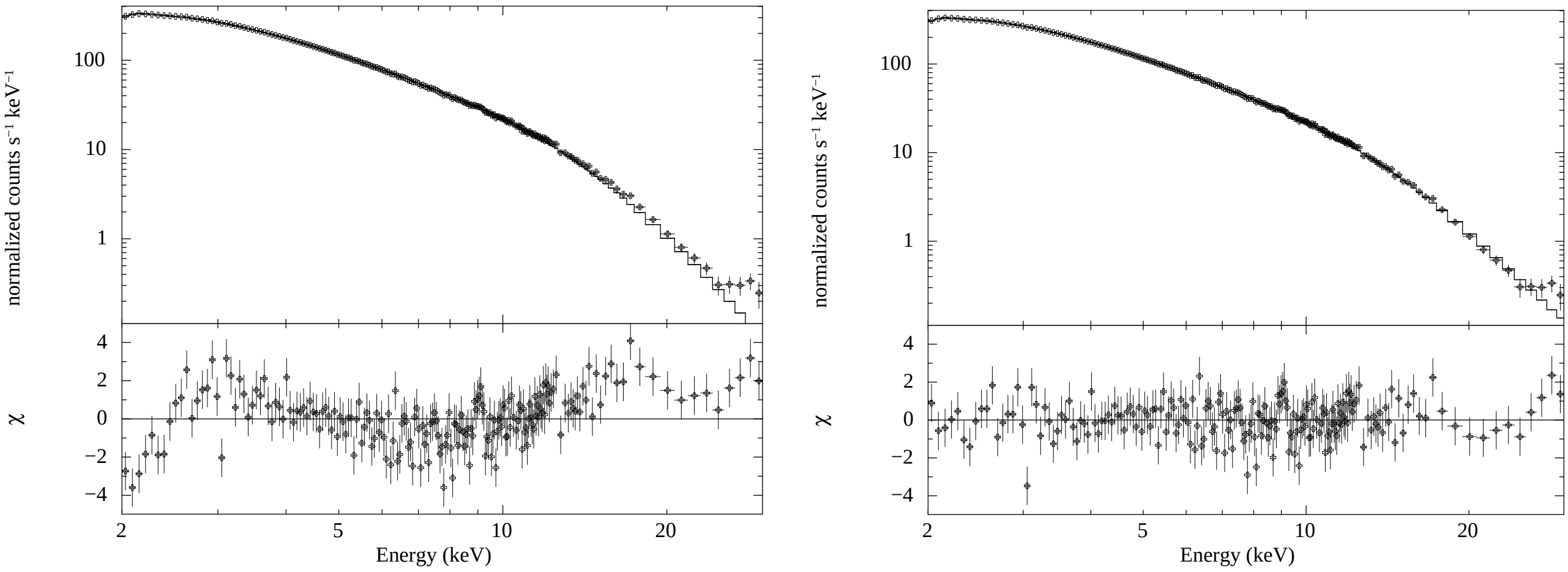

Figure 3: *Left:* LAD: log-parabolic spectrum fitted with a power-law cutoff (top) and the model fit residuals (bottom). *Right:* LAD: same model fitted with a log-parabola.

The LAD will therefore explore the curvature of the X-ray spectrum around 10 keV for HBLs with syncrotron peak at 1–2 keV. Moreover, the capability to extract detailed spectra with a sub-ks temporal integration during the higher states together with the low energy threshold of 2 keV will complement the current understanding provided by X-ray observatories such as NuSTAR (Harrison et al., 2013).

## 2.2 FSRQs: X-$\gamma$ connection and bulk-Comptonization

Contrary to HBLs, FSRQs are blazars whose emission peaks at low energies, with the synchrotron component peaking in the radio/IR bands and the inverse Compton (IC) one in the *Fermi* and AGILE bands (50 MeV–50 GeV). As such, they were not expected to significantly emit at VHE (> 0.2 TeV). The strong and highly variable emission recently discovered by TeV telescopes (e.g., 3C 279, 4C 21.35, PKS 1510−089) came thus as a surprise. The origin is still unclear, but the LAD can provide the answer.

Short time-scale timing studies to compare the synchrotron and IC variability properties as those outlined above for HBLs will be key here. If the emission in the LAD spectral range is dominated by IC emission, simultaneous measurements of the synchrotron emission by ultra-sensitive radio and (sub-)mm instruments such as SKA and ALMA will provide relevant comparisons of the two emission mechanisms. Alternatively, if the strong and highly variable TeV emission is due to a tail of high-energy electrons, their synchrotron radiation should appear in the hard X-ray band as well (like in Intermediate BL Lac or HBL objects), dominating over the IC one and completely changing the spectral slope (from $\beta \sim 1.5$ to $\beta > 2$). Timing is critical: given the highly variable TeV emission (minutes-hours), such X-ray features can disappear quickly and not be measurable by instruments of smaller collecting area or poor pointing flexibility. In this respect, *LOFT*/LAD larger area would provide 1-s binned light curves for flux ≥ 1 mCrab ($\sim 2 \times 10^{-11}$ erg cm$^{-2}$ s$^{-1}$ in 2–10 keV) and, in addition, spectral trend investigation during the flaring activity will be performed on minute time scale, i.e., the non-thermal continuum power-law will be constrained within a few percent.

A further diagnostic will be provided by the LAD on the jet composition, in connection with GeV flares. If the jet is matter-dominated and ejected in blobs, the cold electrons are expected to Compton-upscatter Broad Line Region (BLR) photons yielding a spectral signature as an excess emission at $\sim (\Gamma/10)^2$ keV, where $\Gamma$ is the bulk-motion Lorentz factor of the jet (Sikora & Madejski, 2000). This feature, referred as bulk-Comptonization, has not been seen yet, but 1) with values of $\Gamma > 20$, this feature could actually peak around 10 keV, and thus could not be recognized so far; 2) if the jet accelerates slowly, it could be visible only for few hours, from the time the blob becomes relativistic to the time it goes outside the BLR (at a distance of $\sim 10^{18}$ cm from the central engine) as reported by Celotti et al. (2007) (see their Fig. 1). The LAD will be able to discover this feature for the first time, or put very strong upper limits on the particle content of the jet at its base.

A complementary and independent assessment of the jet composition will be possible with future ultra-sensitive radio instruments such as SKA through modeling of full-polarimetric (linear and circular polarization) spectra (Homan et al., 2009; Agudo et al., 2015).

Concerning intermediate or quiescent flux states, it is generally assumed that FSRQs, and probably also LBLs and IBLs, require target photons from external photon fields (such as BLR, accretion disk, dust torus, host galaxy,. . . ) for an EC component that can account for the high-energy part of their SEDs (e.g., Böettcher et al., 2012). For the brightest of these objects, *LOFT*/LAD could help to distinguish such components from SSC emission by their different temporal behaviour and by comparison with the temporal and spectral properties of the synchrotron spectrum as observed by radio to IR facilities, which would be an important step towards a unified view of different blazar classes.

Combining LAD repointing with radio and (sub-)mm observations with SKA, EVLA, and ALMA can have also a strong impact on the SED modeling and thus on the nature of the $\gamma$-ray emitting region, being the higher radio frequencies (30–950 GHz) properly related to the flaring activity. Further, long term monitoring of FSRQ and LBL-IBLs in all available spectral ranges, including X- and $\gamma$-ray timing on weekly/daily time scales, photo-polarimetry if possible, and millimeter very long baseline interferometric (VLBI) imaging has proven to be a powerful tool to establish the relative location of the different emitting regions from the radio domain up to $\gamma$-rays (Marscher et al., 2010; Agudo et al., 2011). In that regard, WFM will not only provide the triggering of dedicated coordinated observations of particularly interesting events, but also excellent long term X-ray light curves for bright blazars sampled on time scales of up to one day or better.

## 3 The *LOFT* perspective on misaligned AGN

The *LOFT* capabilities can be exploited also to investigate another class of radio-loud sources, the Misaligned AGN (MAGN). With the term MAGN we refer to radio galaxies (radio galaxies have been historically classified as edge-darkened FRI and edge-brightened FRII, on the basis of their extended morphology that changes over, or under, a critical radio power at 178 MHz, $P_{178\,\mathrm{MHz}} \sim 10^{25}\,\mathrm{W\,Hz^{-1}\,s^{-1}}$; Fanaroff & Riley, 1974) and steep spectrum radio quasars, i.e., radio sources with the jet pointed away from the observer's line-of-sight offering a unique opportunity to study the jet phenomenon from a perspective different, but complementary, to that of blazars. MAGN are generally characterized by steep radio spectra ($\alpha_r > 0.5$) and resolved and possibly symmetrical radio structures. Moreover, given the large jet inclination angle the non-thermal radiation produced by MAGN jets is weakly Doppler boosted.

Radio galaxies hosting efficient accretion disks (mainly FRIIs), are unique extragalactic laboratories where a simultaneous view of the accretion flow and ejection of relativistic plasma is possible through multi-wavelength observations. High-resolution radio images, combined with high-energy studies, allow to follow the temporal and spectral evolution of the jet, while X-ray monitoring can provide information on the accretion processes. Up to now there are at least two broad line radio galaxies, i.e., 3C 120 and 3C 111, for which signatures of a possible link between events occurring in the jet and in the accretion disk have been found (Marscher et al., 2002; Chatterjee et al., 2009, 2011) similarly to those observed in X-ray binary systems (Mirabel & Rodriguez, 1998). The disk/jet connection is revealed by a sudden decrease in the X-ray fluxes shortly followed by the appearance of superluminal radio emitting features propagating along the jet. This phenomenon has been interpreted as resulting from an instability in the accretion disk with matter in the inner region partially ejected down the jet (Marscher et al., 2002). Fast disk winds revealed as an absorption blue-shifted feature at around 7 keV, if confirmed, could be also a signature of the ongoing disk/jet perturbation. As this is a random and transient phenomenon, the WFM on-board *LOFT* is the ideal instrument to perform an accurate X-ray monitoring in order to catch the perturbation exactly during its occurrence. The sign of this catastrophic event is a dip in the X-ray light curve. The flux decreases by about a factor of two and the source remains in a low state for few months. The WFM can trace the temporal evolution of powerful radio galaxies with fluxes in the 2–10 keV band above $10^{-11}$ erg cm$^{-2}$ s$^{-1}$ allowing to detect a decrease in the X-ray flux by at least a factor of two on time-scale of months. The WFM can then trigger a LAD pointing of the target to study in detail the spectral evolution of the source during the accretion modification (for example, a suppression of the iron line produced in the inner region of the disk is expected). The WFM dip detection can be used to trigger VLBI observations that can trace the expected ejection of a new emission feature from the radio core. Since the ejections of new emission features are connected to flares at mm-wavelengths (Savolainen et al., 2002), it is also possible to combine the WFM dip monitoring with existing long-term mm-wavelength monitoring programs like Metsähovi 37 GHz monitoring (e.g., Teräsranta et al., 2004) for studying the disk-jet connection. In 3C 120 X-ray flux and 37 GHz flux are indeed anticorrelated with X-rays leading as expected (Chatterjee et al., 2009). This study can be applied at least to other three nearby broad line radio galaxies, i.e. 3C 390.3, 3C 382, 3C 445.

Radio galaxies hosting inefficient accretion flows (mainly FRIs) are more easily detected at high and very high energies. FRIs are more numerous than FRIIs in the GeV sky and four of them have been also detected at TeV energies with current imaging atmospheric Cherenkov telescopes (M87, NGC 1275, Centaurus A, IC 310). Unlike FRIIs, the radio-to-TeV spectrum of low power radio galaxies is dominated by non-thermal emission. Considering that in these sources the Doppler boosting is less important than in blazars, the study of their SED can carry out important information on the jet structure. Indeed, the assumption of a pure, one-zone homogeneous, Synchrotron Self-Compton (SSC) emission region is inadequate (Abdo et al., 2009a,b; Migliori et al., 2011) in these source, while a stratified jet with different regions interacting each other could be a possible solution (Georganopoulos & Kazanas, 2003; Ghisellini et al., 2005; Böttcher et al., 2010). Hadronic models could be another possibility. These have become very popular in the last years because of their direct connection to Ultra High Energy Cosmic Rays (UHECR) and to the recent discovery of high-energy neutrinos by the IceCube detector (Aartsen et al., 2013). Cosmic rays, accelerated at different sites in AGN, could produce $\gamma$-rays and neutrinos (Becker et al., 2009). Among AGNs, radio galaxies are the most prominent candidates (Becker et al., 2014). Estimates of the neutrino flux in the two radio galaxies Centaurus A and M87 have been recently attempted. In order to address this issue the study and modeling of the SED of FRI radio sources (in different luminosity states) is mandatory. *LOFT* is the satellite that can efficiently cover the X-ray region. FRIs are generally less bright than FRIIs between 2–10 keV. However, with an exposure of 10 ks or more the LAD guarantees a $5\sigma$ detection for sources as weak as a few $10^{-12}$ erg cm$^{-2}$ s$^{-1}$. Note that, three out of four TeV radio galaxies have fluxes greater than $10^{-12}$ erg cm$^{-2}$ s$^{-1}$. *LOFT* can be then the reference X-ray satellite for future multi-wavelength campaigns on FRIs led by the CTA observatory monitoring preferentially quiescent states. At the same time, if a CTA alert is issued, the *LOFT* capability to rapidly point the flaring target, will offer the unique opportunity to follow the spectral/flux evolution of a radio galaxy in a high luminosity state.

## 4 The *LOFT* perspective on radio-loud narrow line Seyfert 1 galaxies

The discovery by *Fermi*-LAT of variable $\gamma$-ray emission from a few radio-loud narrow-line Seyfert 1s (NLSy1s) revealed the presence of a possible third class of AGN with relativistic jets (Abdo et al., 2009; D'Ammando et al., 2012). By considering that NLSy1s are usually thought to be hosted in spiral galaxies (e.g., Deo et al., 2006), the presence of a relativistic jet in these sources seems to be in contrast to the paradigm that the formation of those structures could happen only in elliptical galaxies (e.g., Marscher, 2009). Moreover, smaller masses of the central BH and higher accretion rates than those observed in blazars and radio galaxies are inferred for NLSy1s.

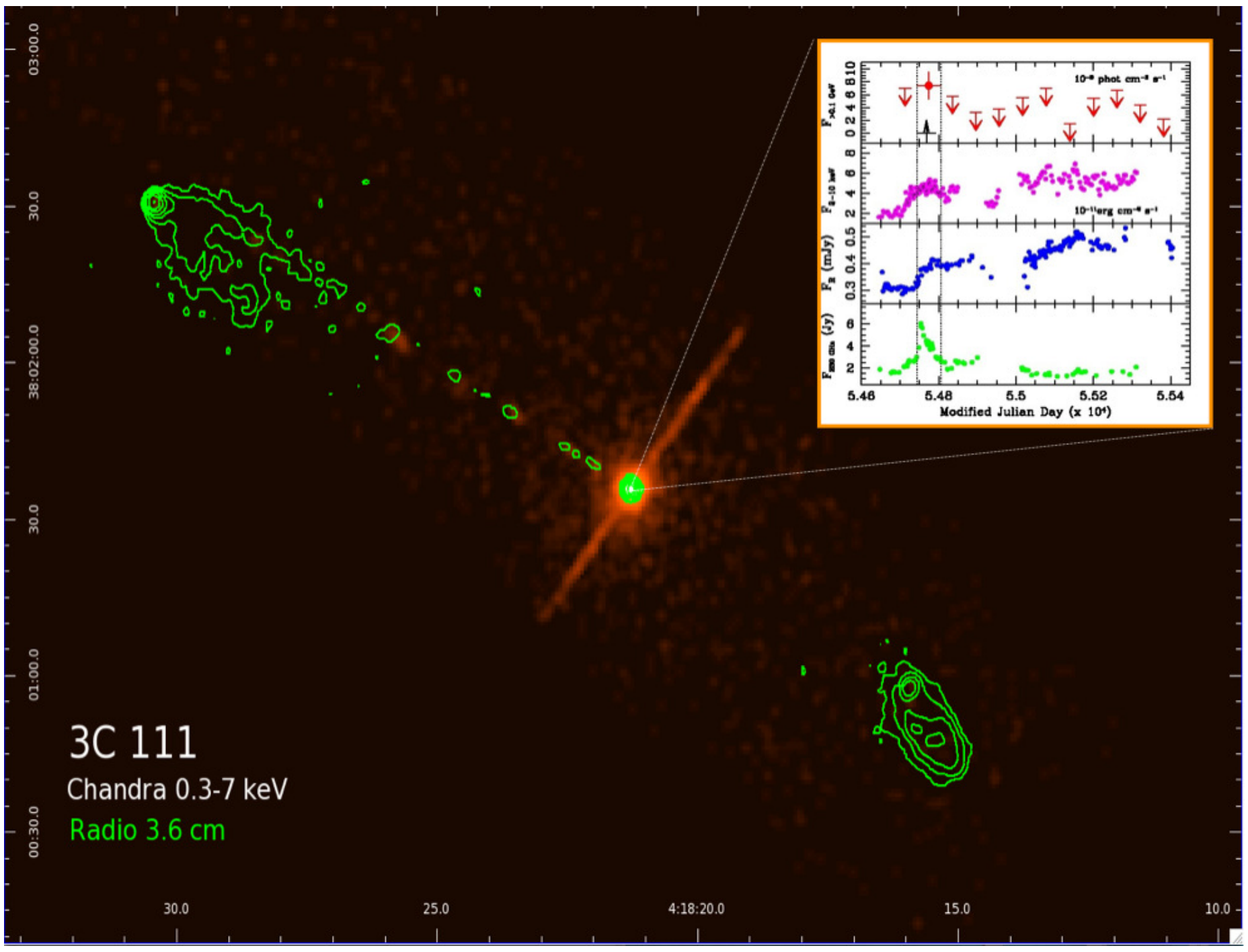

Figure 4: Chandra 0.3–7 keV image of the FRII radio galaxy 3C 111 with 3.6 cm radio contours overimposed. In the inset the multi-wavelength light curve from mm to $\gamma$-rays is shown. The simultaneity of the flare is impressive: the core luminosity was increasing from millimeter to X-ray frequencies exactly when the greatest flux of $\gamma$-ray photons occurred. This is a clear indication of the cospatiality of the event. The outburst of photons is directly connected to the ejection of a new radio knot (the time of the ejection is indicated by the black arrow).

In this context the detection of NLSy1s in $\gamma$-rays poses intriguing questions about the nature of these objects, the onset of production of relativistic jets, the mechanisms of high-energy production, and the jet-disc connection.

Strong $\gamma$-ray flares were observed from PMN J0948+0022 (Foschini et al., 2011; D'Ammando et al., 2015), SBS 0846+513 (D'Ammando et al., 2013), and 1H 0323+342 (Carpenter et al., 2013), with an isotropic $\gamma$-ray luminosity of $10^{48}$ erg s$^{-1}$, comparable to that of the bright FSRQs, and flux variability on a daily time scale or shorter. Follow-up observations of NLSy1s during flaring activity with the *LOFT*/LAD will be important for investigating rapid flux and spectral changes occurring in X-rays. This will be fundamental for constraining size and location of the high-energy emitting region (also in conjunction with CTA) and the possible presence during those high activity states of two different emitting components (e.g. SSC, external Compton from BLR and/or dust torus) in the broad energy range covered by LAD, characterizing in detail the SED of these sources.

Furthermore, 1H 0323+342 is the brightest NLSy1 (1–2 $\times 10^{-11}$ erg cm$^{-2}$ s$^{-1}$ in 2–10 keV) in X-rays among the 5 sources detected by *Fermi*-LAT so far, and the only one included in the 70-month *Swift*-BAT catalogue as well as significantly detected by INTEGRAL (Baumgartner et al., 2013; Panessa et al., 2011). Therefore it is the ideal target for investigating the disc-jet connection in the $\gamma$-ray NLSy1. The sum of the X-ray spectra of 1H 0323+342 collected by *Swift*-XRT showed the hint for the presence of an iron K$\alpha$ line (Paliya et al., 2014). Thanks to the high-quality spectrum collected by the LAD we will be able to confirm the detection of the iron K$\alpha$ line, and test with high accuracy the reflection model applied to the X-ray spectrum thanks to its broader energy range, thus removing the degeneracy of the parameters. This will allow us to constrain the properties of the accretion flow. In particular, we will be able to obtain very good constraints on the spin of the black hole. This information can be used for probing the inner accretion disc, a region strictly related to the jet formation and launching. If the high spin of the black hole claimed by Paliya et al. (2014) is confirmed by the high-quality

LAD data, this will be another important indication that high spin is necessary for the jet production.

## 5 Conclusions

We discussed the potential of *LOFT* for the study of radio-loud AGNs mainly based on its timing capabilities exploited in synergy with other observatories (radio, TeV) planned to operate in 2020s. Although blazars seem to be the most obvious candidates because of their extreme variability, there will be a big space for investigation of the emitting properties of several classes of radio-loud AGNs.

On the basis of the portion of the spectral energy distribution accessible to *LOFT* (2-50 keV), we argue that among blazars, HBL objects are the best candidates for pointed observations with LAD. These observations will provide us the best sampling of the light curves in flaring states, given the exploration of unprecedented short timescales comparable with those achieved by the future TeV observatory CTA and in strict simultaneity. This will be possible for a large sample of TeV blazars thanks to the LAD pointing flexibility (with ~ 75% of the sky accessible). Therefore, *LOFT* will open a new window on the investigation of X-ray/TeV connection, with particular regard to:

- lag measurements with at least an order of magnitude improvement over the current limit of ~100 seconds;
- the study of the multi-zone SSC role during flares;
- the connection between the jet variability and the central engine.

Moreover, the capability to extract detailed 2–50 keV spectra with a sub-ks temporal integration (during the higher states) will allow us to study the mechanism of acceleration of highly energetic electrons, by following the curvature variations of the synchrotron emission.

Although FSRQs are less bright and variable in X-rays (see, e.g., Bhattacharyaet al., 2013) with respect to HBLs, LAD follow-up observations will allow us to investigate temporal (second timescale) and spectral variations during typical flaring activity (minute timescale). In a multi-frequency context, *LOFT* will contribute to identify the nature and location of the high energy emission (particularly if detected at TeV energies which explore similar timescales).

It should also be noted that there is a great potential of simultaneous flare coverage with CTA and *LOFT*/LAD for non-blazar AGN, especially gamma-loud radio galaxies. Past experience with multi-wavelength campaigns on flares of M87 for example (Abramowski et al., 2012) have proven extremely valuable in constraining the location of high-energy emission in this radio galaxy. New scenarios open up from the recent detection of a fast TeV-flare from AGN IC 310, an intermediate object between radio galaxies and blazars (Aleksic et al., 2014). The minute-scale doubling time measured in the light curve, the small Doppler factor and the energetics of the flare pose new challenges on both the acceleration mechanism and the $\gamma$-ray emitting zone, presumably a very compact region located at the magnetosphere surrounding the black hole. The possibility to discriminate among different acceleration processes is strictly connected to the possibility of a simultaneous detection on other wavelengths of such extreme activity. *LOFT*/LAD will be the instrument capable of providing indications of X-ray variability on time scales of few tenths of seconds at the flux level of $10^{-11}$ erg cm$^{-2}$ s$^{-1}$ from IC 310 and similarly from new intermediate objects that will emerge in the TeV sky thanks to the increased sensitivity of CTA. LAD repointing of the high energy flares from NLSy1 detected in $\gamma$-rays or at higher energy $\gamma$-rays will allow to follow the X-ray spectral variations of these sources down to a few $10^{-12}$ erg cm$^{-2}$ s$^{-1}$, unveiling the interplay between the comptonized radiation from corona and the non-thermal emission from the jet.

Finally, the sky coverage, the 1-day sensitivity (~ a few mCrab at $5\sigma$ for an on-axis source) and the pointing flexibility of the WFM as well will provide triggers for multiwavelength follow-up from radio (SKA, ALMA) up to TeV energies (CTA), crucial for both blazars and radio galaxies sources. WFM will also provide much needed long-term monitoring of X-ray bright radio galaxies and blazars on time scales of weeks and months.

Table 1: The sample of AGNs detectable with the LAD

| source | $\alpha$ (J2000) | $\delta$ (J2000) | Flux (2–10 keV) [erg cm$^{-2}$ s$^{-1}$] | class |
|---|---|---|---|---|
| S5 0014+81 | 4.2853 | 81.5855 | $9.9\times10^{-11}-1.5\times10^{-10}$ | FSRQ |
| 1ES 0033+595 | 8.9694 | 59.8346 | $5.9\times10^{-11}$ | HBL |
| RBS 97 | 10.4445 | −47.0281 | $8.693\times10^{-12}$ | BLLac |
| 5C 3.178 | 11.9821 | 39.8185 | $6.289\times10^{-12}$ | BLLac |
| (HB89) 0110+318 | 18.2110 | 32.1363 | $5.245\times10^{-12}$ | FSRQ |
| 1ES 0120+340 | 20.7860 | 34.3468 | $1.3\times10^{-11}-1.7\times10^{-11}$ | BLLac (HBL) |
| GB6 J0136+3906 | 24.1354 | 39.0998 | $1.1\times10^{-11}$ | BLLac (HBL) |
| PKS 0208−512 | 32.6925 | −51.0171 | $7.6\times10^{-11}$ | HP-BLLAC |
| 4C +06.11 | 36.1175 | 6.9890 | $6.502\times10^{-12}$ | FSRQ |
| BWE 0229+2004 | 38.2019 | 20.2888 | $1.687\times10^{-11}$ | BLLac (HBL) |
| IC 310 | 49.1792 | 41.3247 | $2\times10^{-12}$ | FRI/BLLAC |
| RBS 0413 | 49.9658 | 18.7596 | $7.0\times10^{-12}-1.5\times10^{-11}$ | BL Lac (HBL) |
| NGC 1275 | 49.9507 | 41.5117 | $4.3\times10^{-11}-4.1\times10^{-10}$ | FRI |
| 1H 0323+342 | 51.1715 | 34.1794 | $1\times10^{-11}-2\times10^{-11}$ | NRLSy 1 |
| PKS 0332−403 | 53.5568 | −40.1403 | $1.2\times10^{-10}$ | HP BLLac |
| NRAO 140 | 54.1254 | 32.3081 | $7.2\times10^{-12}$ | FSRQ |
| 1ES 0414+009 | 64.2187 | 1.0899 | $7.8\times10^{-12}-1.0\times10^{-11}$ | BL Lac (HBL) |
| PKS 0420−01 | 65.8158 | −1.3425 | $1.6\times10^{-11}$ | FSRQ |
| 3C 120 | 68.2962 | 5.3543 | $4\times10^{-11}$ | FRII |
| PKS 0447−439 | 72.3529 | −43.8358 | $5.0\times10^{-12}$ | BL Lac (HBL) |
| 1ES 0507−040 | 77.4091 | −4.0126 | $5.5\times10^{-12}$ | BL Lac (HBL) |
| 1ES 0502+675 | 76.9848 | 67.6234 | $3.9\times10^{-11}$ | BL Lac (HBL) |
| PKS 0521−36 | 80.7416 | −36.45856 | $8.7\times10^{-12}-1.4\times10^{-11}$ | FSRQ |
| PKS 0528+134 | 82.7350 | 13.5320 | $9.0\times10^{-12}$ | FSRQ |
| PKS 0537−158 | 84.8807 | −15.8431 | $1.287\times10^{-11}$ | FSRQ |
| PMN J0546−6415 | 86.6743 | −64.2561 | $8.2\times10^{-12}$ | FSRQ |
| PKS 0548−322 | 87.6694 | −32.2714 | $1.6\times10^{-11}-3.0\times10^{-11}$ | BL Lac (HBL) |
| RX J0648.7+1516 | 102.1985 | 15.2735 | $5.0\times10^{-12}$ | BL Lac (HBL) |
| 1ES 0647+250 | 102.6937 | 25.0499 | $2.3\times10^{-11}$ | BLLac (HBL) |
| RGB J0710+591 | 107.6283 | 59.1352 | $2.290\times10^{-11}$ | BLLac (HBL) |
| S5 0716+714 | 110.4727 | 71.3434 | $5.0\times10^{-12}$ | BL Lac (IBL) |
| 2MASX J07315268+2804333 | 112.9709 | 28.0762 | $1.288\times10^{-11}$ | BL Lac |
| PKS 0743−67 | 115.8763 | −67.4413 | $1.157\times10^{-11}$ | FSRQ |
| PKS 0748+126 | 117.7165 | 12.5170 | $5.200\times10^{-12}$ | FSRQ |
| 1ES 0806+524 | 122.4549 | 52.3162 | $8.7\times10^{-12}-1.3\times10^{-11}$ | BL Lac (HBL) |
| PMN J0816−1311 | 124.1133 | −13.1980 | $5.7\times10^{-12}$ | BLLac |
| GB6 J0827+5218 | 126.9739 | 52.3000 | $5.758\times10^{-12}$ | FSRQ |
| 1ES 0836+710 | 130.3499 | 70.8948 | $2.894\times10^{-11}$ | FSRQ |
| RBS 0723 | 131.8039 | 11.5640 | $5.0\times10^{-12}$ | BL Lac (HBL) |
| SDSS J090816.77+052011.9 | 137.0707 | 5.3366 | $5.719\times10^{-12}$ | QSO |
| PKS 0921−213 | 140.9120 | −21.5964 | $5.0\times10^{-12}-13.9\times10^{-12}$ | FSRQ/Sy1 |
| 1ES 0927+500 | 142.6566 | 49.8404 | $6.6\times10^{-12}$ | BLLac (HBL) |
| NVSS J101015−311906 | 152.5665 | −31.3191 | $7.5\times10^{-12}$ | BLLac |
| 1ES 1011+496 | 153.7672 | 49.4335 | $1.0\times10^{-11}$ | BL Lac (HBL) |

*continued on next page*

*continued from previous page*

| source | $\alpha$ (J2000) | $\delta$ (J2000) | Flux (2–10 keV) [erg cm$^{-2}$ s$^{-1}$] | class |
|---|---|---|---|---|
| 1ES 1028+511 | 157.8272 | 50.8933 | $7.7\times10^{-12}$ – $1.0\times10^{-11}$ | BL Lac (HBL) |
| Mrk 421 | 166.1138 | 38.2088 | $6.3\times10^{-11}$ - $6.2\times10^{-10}$ | BL Lac |
| RBS 0970 | 170.2014 | 42.2033 | $1.564\times10^{-11}$ | BLLac |
| PKS 1127−14 | 172.5293 | −14.8243 | $5.3\times10^{-12}$ – $7.4\times10^{-12}$ | FSRQ |
| SDSS J113025.21+400446.3 | 172.6068 | 40.0798 | $1.021\times10^{-11}$ | QSO |
| Mrk 180 | 174.1100 | 70.1576 | $5.0\times10^{-12}$ – $1.0\times10^{-11}$ | BL Lac (HBL) |
| RX J1136.5+6737 | 174.1253 | 67.6178 | $5.0\times10^{-12}$ | BL Lac (HBL) |
| PKS 1145−676 | 176.9071 | −67.8902 | $8.845\times10^{-12}$ | FSRQ |
| RX J1211.9+2242 | 182.9943 | 22.7091 | $5.2\times10^{-12}$ – $7.5\times10^{-12}$ | BL Lac (HBL) |
| 1ES 1218+304 | 185.3414 | 30.1770 | $1.5\times10^{-11}$ | BL Lac (HBL) |
| 3C 273 | 187.2779 | 2.0523 | $5.9\times10^{-11}$ – $1.9\times10^{-10}$ | FSRQ-Sy1.0 |
| M 87 | 187.7059 | 12.3911 | $1.9\times10^{-11}$ | FRI |
| PMN J1253−3932 | 193.4223 | −39.5322 | $7.5\times10^{-12}$ – $8.45\times10^{-12}$ | BL Lac |
| SDSS J125436.40+053720.4 | 193.6515 | 5.6219 | $5.242\times10^{-12}$ | QSO |
| 3C 279 | 194.0451 | −5.7880 | $1.4\times10^{-11}$ – $2.65\times10^{-11}$ | FSRQ |
| 1ES 1255+244 | 194.3831 | 24.2111 | $1.2\times10^{-11}$ | BL Lac (HBL) |
| SDSS J132144.96+033055.7 | 200.4389 | 3.5150 | $7.414\times10^{-12}$ | QSO |
| Centaurus A | 201.3651 | −43.0191 | $2.0\times10^{-11}$ – $9.2\times10^{-11}$ | FRI |
| PMN J1337−1257 | 204.4157 | −12.9568 | $6.0\times10^{-12}$ – $9.6\times10^{-12}$ | FSRQ |
| (DGT2001) J1410+6100 | 212.6265 | 61.0030 | $5.754\times10^{-12}$ | BLLac |
| SDSS J141758.60+091609.7 | 214.4934 | 9.2702 | $6.689\times10^{-12}$ | QSO? |
| RBS 1366 | 214.4861 | 25.7240 | $8.8\times10^{-12}$ – $1.2\times10^{-11}$ | BL Lac (HBL) |
| PKS 1424+420 | 216.7516 | 23.8000 | $5.0\times10^{-12}$ | BL Lac (HBL) |
| H 1426+428 | 217.1361 | 42.6724 | $1.8\times10^{-11}$ – 3.6E-11 | BL Lac (HBL) |
| PKS 1440−389 | 220.9882 | −39.1442 | $5.1\times10^{-12}$ | BLLac |
| RBS 1434 | 222.3855 | 27.7734 | $1.008\times10^{-11}$ | BLLac |
| RBS 1457 | 225.9195 | −15.6872 | $5.5\times10^{-12}$ | BLLac |
| PKS 1510−08 | 228.2077 | −9.1020 | $1.003\times10^{-11}$ | FSRQ |
| PKS 1514−24 | 229.4242 | −24.3720 | $5.1\times10^{-12}$ – $8.1\times10^{-12}$ | BLLac |
| (DGT2001) B1533+535 | 233.7580 | 53.3430 | $8.715\times10^{-12}$ | BLLac |
| RX J1548.3+6949 | 237.0655 | 69.8284 | $6.357\times10^{-12}$ | QSO |
| PMN J1548−2251 | 237.2072 | −22.8507 | $6.9\times10^{-12}$ | BLLac |
| PG 1553+113 | 238.9294 | 11.1901 | $1.4\times10^{-11}$ – $3.5\times10^{-11}$ | BL Lac (HBL) |
| PKS 1549−79 | 239.2453 | −79.2345 | $5.2\times10^{-12}$ – $5.8\times10^{-12}$ | FSRQ/Sy1i |
| PKS 1610−77 | 244.4700 | −77.2898 | $6.273\times10^{-12}$ | FSRQ |
| 87GB 162418.8+435342 | 246.4692 | 43.7864 | $7.190\times10^{-12}$ | FSRQ |
| PKS 1622−29 | 246.5251 | −29.8574 | $5.0\times10^{-12}$ – $6.0\times10^{-12}$ | FSRQ |
| PKS 1635−14 | 249.6875 | −14.2654 | $6.243\times10^{-12}$ | FSRQ? |
| 4C 39.48 | 250.7450 | 39.8086 | $5.1\times10^{-12}$ – $6.6\times10^{-12}$ | FSRQ |
| Mrk 501 | 253.4675 | 39.7602 | $4.5\times10^{-11}$ – $21.9\times10^{-11}$ | BL Lac |
| Swift J1656.2−3301 | 254.0699 | −33.0368 | $5.5\times10^{-12}$ – $6.0\times10^{-12}$ | FSRQ |
| 1H 1720+117 | 261.2685 | 11.8704 | $1.564\times10^{-11}$ | BLLac |
| PKS 1730−13 | 262.0829 | −14.2655 | $6.7\times10^{-12}$ | FSRQ |
| 1E 1727+502 | 262.0776 | 50.21957 | $6.6\times10^{-12}$ | BLLac (HBL) |
| 4C +51.37 | 265.1547 | 52.1943 | $5.629\times10^{-12}$ | FSRQ |
| (DGT2001) B1741+196 | 265.9910 | 19.5863 | $6.9\times10^{-12}$ – $2.2\times10^{-11}$ | BLLac (HBL) |
| PKS 1830−21 | 278.4163 | −21.0610 | $1.1\times10^{-11}$ – $1.5\times10^{-11}$ | FSRQ-LensedQ |

*continued on next page*

*continued from previous page*

| source | $\alpha$ (J2000) | $\delta$ (J2000) | Flux (2–10 keV) [erg cm$^{-2}$ s$^{-1}$] | class |
|---|---|---|---|---|
| 3C 382 | 278.7641 | 32.6963 | $3\times10^{-11} - 6\times10^{-11}$ | FRII |
| ESO 045- G 011 | 280.1654 | −77.1576 | $5.9\times10^{-12} - 1.394\times10^{-11}$ | blazar/SO-PEC |
| 3C 390.3 | 280.5374 | 79.7714 | $4\times10^{-11}$ | FRII |
| PKS 1914−45 | 289.4148 | −45.5083 | $7.827\times10^{-12}$ | QSO |
| PKS B1921−293 | 291.2127 | −29.2417 | $5.3\times10^{-12} - 9.8\times10^{-12}$ | FSRQ |
| NVSS J193347+325426 | 293.4487 | 32.9069 | $1.0\times10^{-11} - 1.3\times10^{-11}$ | QSO |
| PMN J1936−4719 | 294.2337 | −47.3305 | $7.8\times10^{-12}$ | BLLac |
| 1ES 1959+650 | 299.9994 | 65.1485 | $1.4\times10^{-11} - 2.4\times10^{-10}$ | BLLac (HBL) |
| PMN J2009−4849 | 302.3558 | −48.8315 | $5.3\times10^{-12} - 5.9\times10^{-11}$ | BLLac (HBL) |
| 4C +21.55 | 308.3838 | 21.7727 | $1.575\times10^{-11}$ | FSRQ |
| PKS 2052−47 | 314.0709 | −47.2471 | $8.578\times10^{-12}$ | FSRQ |
| GB6 J2109+3532 | 317.3802 | 35.5495 | $1.185\times10^{-11}$ | FSRQ |
| S5 2116+81 | 318.5049 | 82.0801 | $1.2\times10^{-11} - 1.6\times10^{-11}$ | FSRQ |
| PKS 2126−15 | 322.3010 | −15.6467 | $5.8\times10^{-12} - 2.5\times10^{-11}$ | FSRQ |
| RBS 1787 | 327.5645 | −14.1806 | $6.3\times10^{-12} - 8.8\times10^{-12}$ | BLLac |
| PKS 2149−306 | 327.9806 | −30.4652 | $7.6\times10^{-12} - 13.7\times10^{-12}$ | FSRQ |
| PKS 2155−304 | 329.7169 | −30.2255 | $2.3\times10^{-11} - 8.8\times10^{-11}$ | BL Lac (HBL) |
| BL Lac | 330.6804 | 42.2778 | $5.8\times10^{-12} - 1.3\times10^{-11}$ | BL Lac (IBL) |
| 3C 445 | 335.9570 | -2.1036 | $2\times10^{-11}$ FRII | |
| PKS 2227−08 | 337.4174 | −8.5482 | $9.417\times10^{-12}$ | FSRQ |
| PMN J2230−3942 | 337.6678 | −39.7145 | $5.2\times10^{-12}$ | FSRQ |
| PKS 2230+11 | 338.1517 | 11.7308 | $5.5\times10^{-12} - 9.8\times10^{-12}$ | FSRQ |
| PKS 2233-148 | 339.1418 | −14.5563 | $5.6\times10^{-12} - 6.2\times10^{-12}$ | BLLac |
| RBS 1895 | 341.6739 | −52.1103 | $9.640\times10^{-12}$ | BLLac |
| RBS 1906 | 342.9457 | −32.1036 | $1.267\times10^{-11}$ | BLLac |
| 3C 454.3 | 343.4906 | 16.1482 | $3.4\times10^{-11} - 1.3\times10^{-10}$ | FSRQ |
| HE 2327−5522 | 352.5047 | −55.1096 | $7.172\times10^{-12}$ | QSO |
| PKS 2331−240 | 353.4801 | −23.7279 | $5.7\times10^{-12} - 8.1\times10^{-12}$ | FSRQ/Sy? |
| SHBLJ234333.8+344004 | 355.8914 | 34.6652 | $9.288\times10^{-12}$ | BLLac |
| 1ES 2344+514 | 356.7702 | 51.70497 | $8.5\times10^{-12} - 6.3\times10^{-11}$ | BLLac (HBL) |
| PKS 2345−16 | 357.0102 | −16.5222 | $7.495\times10^{-12}$ | FSRQ |
| RBS 2066 | 359.3750 | −17.3013 | $6.0\times10^{-12}$ | BLLac |
| H 2356−309 | 359.7830 | −30.6279 | $9.1\times10^{-12} - 5.3\times10^{-11}$ | BLLac (HBL) |

## References

Aartsen, M.G. (IceCube Coll.), et al. 2013, Science, 342

Abbasi, R. U., Abe, M., Abu-Zayyad, T., et al., ApJ, 790, L21 (2014)

Abdo, A. A., Ackermann, M., Ajello, M., et al. 2009a, ApJ, 706, 275

Abdo, A. A., Ackermann, M., Ajello, M., et al. 2009b, ApJ, 707, 55

Abdo, A. A., et al. 2009c, ApJ, 707, L142

Abramowski, A. et al., ApJ, 746, 151 (2012)

Agudo, I., Marscher, A. P., Jorstad, S. G., et al. 2011, ApJ, 735, L10

Agudo, I. et al. 2015, Proc. "Advancing Astrophysics with the Square Kilometre Array", PoS(AASKA14)093, in press

Aharonian, F. et al. 2009, A&A, 502, 749

Aleksic, J. et al. 2014, Science, 346, 1080

Baumgartner, W. H., et al. 2013, ApJ Suppl., 207, 19

Becker, J.K. and Biermann, P.L., 2009, Astropart. Phys., 31, 138

Becker Tjus, J. et al. 2014, Physical Review D, 89, 123005

Begelman, M.. C., Fabian, A. C., and Rees, M. J. 2008 MNRAS, 384, L19

Bhattacharya, D., Misra, R., Rao, A. R., and Sreekumar, P. et al. 2013, MNRAS, 431,1618

Blandford, R. D., & Rees, M. J. 1978, in BL Lac Objects ed A. M. Wolfe (Univ. Pittsburgh Press), 328

Böttcher, M. & Dermer, C. D. 2010, ApJ, 711, 445

Böttcher, M., Reimer, A., Sweeney, K. and Prakash, A. 2012 ApJ, 768, 54

Carpenter, B., & Ohja, R. 2013, ATEL 5344

Chatterjee, R. et al. 2009, ApJ 704, 1689

Chatterjee, R. et al. 2011, ApJ 734, 43

Celotti, A., Ghisellini, G., and Fabian, A. C 2007, MNRAS, 375, 417
Cerruti, M., Zech, A., Boisson, C. and Inoue, S. 2014, arXiv:1411.5968.
D'Ammando, F., et al. 2012, MNRAS, 426, 317
D'Ammando, F., et al. 2013, MNRAS, 436, 191
D'Ammando, F., et al. 2015, MNRAS, 446, 2456
Deo, R. P., et al. 2006, AJ, 132, 321
Fanaroff, B.L. & Riley, J.M. 1974, MNRAS, 167, 31
Feng, K. et al., 2014, ApJ 794, 126
Feroci M., et al. 2014, SPIE, 9144, 91442
Foschini, L., et al. 2011, MNRAS, 413, 1671
Fossati, G., Maraschi, L., Celotti, A., Comastri, A. and Ghisellini, G. 1998, MNRAS, 299, 433
Georganopoulos, M., & Kazanas, D. 2003, ApJ, 589, L5
Ghisellini, G., Tavecchio, F., & Chiaberge, M. 2005, A&A, 432, 401
Grandi, P., Torresi, E. & Stanghellini, C. 2012, ApJ, 751, 3
Ghisellini, G. and Tavecchio, F. 2009, MNRAS, 397, 985
Ghisellini, G., Tavecchio, F., Bodo, G. and Celotti, A. 2009, MNRAS 393, L16
Giannios, D. 2013 MNRAS, 431, 355
Harrison, F. A. et al. 2013 ApJ, 770, 103
Homan, D., et al. 2009, ApJ, 696, 328
Marscher et al. 2002, Nature, 417, 625
Marscher, A. 2009, in Lecture Notes in Physics 794, ed. T. Belloni (Berlin:Springer), 173 [arXiv:0909.2576]
Marscher, A. P., Jorstad, S.G., Larionov, V. M., et al. 2010, ApJ, 710, L126
Marscher, A. 2014, ApJ, 780, 87
Migliori, G., Grandi, P., Torresi, E., et al. 2011, A&A, 533, 72
Mirabel & Rodriguez, 1998, Nature, 392, 673
Narayan, R. and Piran, T. 2012, MNRAS, 420, 604
Paliya, V. S., et al. 2014, ApJ, 789, 143
Panessa, F., et al. 2011, MNRAS, 417, 2426
Rieger, F. M. and Aharonian, F. A. 2008, A&A, 479, L5
Savolainen, T. et al. 2002, A&A, 394, 851
Sikora, M. and Madejski, G. 2000, ApJ, 534, 109
Sokolov, A., Marscher, A. P. and McHardy, I. M. 2004, ApJ, 613, 725
Sol, H. et al. 2013, Astroparticle Physics, 43, 215
Tavecchio, F., Ghisellini, G., & Guetta, D., 2014, ApJ, 793, L18 (2014)
Teräsranta, H. et al., 2004, A&A, 427, 769
Tramacere, A., Massaro, E. and Taylor, A. M. 2011, ApJ, 739, 66
Tramacere, A., Massaro, F. and Cavaliere, A. 2007, A&A, 466, 521
Urry, C. M. and Padovani, P. 1995, PASP, 107, 803